\documentclass{article}

\usepackage[english]{babel}

\usepackage[letterpaper,top=2cm,bottom=2cm,left=3cm,right=3cm,marginparwidth=1.75cm]{geometry}

\usepackage{amsmath}
\usepackage{graphicx}
\usepackage[colorlinks=true, allcolors=blue]{hyperref}
\usepackage{subcaption}

\title{Slip length estimation for flow over lubricant-impregnated surface}
\author{Vishal Goyal, Subhra Datta}

\begin{document}
\maketitle

\begin{abstract}
Lubricant-impregnated surfaces (LIS) and superhydrophobic surfaces (SHSs) are known to passively reduce drag over a surface, which, with a suitable design such as the ribbed texture, can also steer flows anisotropically. Analytical predictions are developed for ribbed textures using an eigenfunction expansion approach. Compared to currently available analytical predictions, these predictions demonstrate superior numerical accuracy and avoid restrictive assumptions on rib geometry, working fluid and lubricant properties.  The predictions provide contrasting design prescriptions depending on whether a lower drag or a larger degree of anisotropic flow deflection is desired. 


\end{abstract}

\section{Introduction}
Any solid body moving through fluids attracts drag forces, which need to be compensated through an external energy supply. Reduction in drag force could make transport of goods through marine vessels \cite{wan2018decarbonizing} and through pipes networks \cite{choquette2018coptem, kyung2017estimation} more efficient. 
Surface iblets have proven to be significant among passive drag reduction methods. 
The fluid flowing over the ribbed surface (referred to as the main fluid) is unable to penetrate the gap between the adjacent ribs for micro and nano-sized riblets. This happens because of the dominant surface-tension force at the small riblet length scales.
Such surfaces having air trapped between the riblet gaps are known as superhydrophobic surfaces (SHS).
However, these surfaces are fragile \cite{bobji2009underwater, varanasi2010frost, samaha2012influence, samaha2012sustainability}, and their durability can be increased by filling the riblets with another fluid, known as a lubricating fluid. Lubricating fluids must be immiscible with the main fluids. Such surfaces are known as lubricant-impregnated surfaces (LIS). In this way, superhydrophobic surfaces are just a special case of LIS, where lubricating fluid is air.

Drag reduction capabilities of the surface are quantified with the parameter known as slip length, which is first defined by Navier \cite{navier1822memoire} as the ratio of fluid's velocity at the reference plane to its gradient, as given below:
\begin{equation}
    b = \dfrac{u}{du/dy}    \label{eq:slipLength}
\end{equation}
where \(u_s\) is the velocity, and \(du/dy\) is its gradient, both measured at the plane touching the peaks of the ribs, which is considered the reference plane for LIS \cite{schonecker2015assessment}.
The slip length suggested by Navier \cite{navier1822memoire} is in context to the intrinsic slip, which is negligible for the channel of micron or larger length scales. The slip provided by the LIS is extrinsic slip, for which the surface average values for the quantities in Eq. \eqref{eq:slipLength} must be used \cite{bazant2008tensorial}.

Changes in local flow field due to ribbed surface were studied using PIV \cite{ou2004laminar, ou2005direct, ou2007enhanced} and numerical simulations \cite{cheng2009microchannel, ge2018effective, maynes2007laminar} in detail. These studies helped in understanding how the ribs affect the flow compared to flow over a flat surface. However, having an analytical solution is necessary for the efficient design of ribs.
Even-though analytical solution over ribbed surface is not easy because of complex surface shape, still it is present for the case of single fluid system \cite{wang2003flow, stroock2002patterning, dewangan2020effective}, where the main fluid exists in Wenzel state and lubricating fluid is absent.
Infusion of lubricating fluid in LIS compounded the complexity further.

The earliest simplification for such a complex problem can be traced to the study of Philip \cite{philip1972flows, philip1972integral}, who employed the shear-free assumption, to estimate the slip length provided by air-filled topographies. 
This approach treats the interface between the main fluid and the lubricant as an ideal slipping surface, offering infinite slip.  Consequently, the two-layer fluid flow problem simplifies into an analysis of  the flow of a single fluid over a flat surface with variable slip (no slip on solid-fluid and infinite slip on fluid-fluid interfaces). This model was later used for analyzing slip length in various configurations such as pipe flow \cite{lauga2003effective}, parallel plate channels \cite{teo2009analysis}, curved interfaces \cite{sbragaglia2007note} and partially filled \cite{ng2009stokes} cavities.

The assumption of a shear-free interface neglects the viscosity of the lubricant and anything beneath the fluid-fluid interface. 
As the viscosity of air is much less than the viscosity of water (the most obvious choice for the main fluid), this assumption gave approximate results for the SHS, but for LIS, where the lubricant viscosity cannot be neglected, this theory became irrelevant. However, later studies using numerical simulations show the discrepancies are even significant for SHS \cite{davies2006laminar, maynes2007laminar, cheng2009microchannel}.
Simulation by Cheng et al. \cite{cheng2009microchannel} shows that this assumption predicts well for the lubricant having viscosity 1000 times lower than the main fluid, while the actual viscosity of air is only about 50 times lower than the water. Hence, this theory has significant scope for improvement.

Within the boundaries of modeling single-phase flows, the gas-cushion theory (GCT) proposed by \cite{vinogradova1995drainage} is a notable improvement over the assumption of the shear-free interface. This theory states that the local slip length at the reference plane is equal to the product of the viscosity ratio of the main fluid to lubricant and the local distance of the plane from the surface, which can be represented as:
\begin{equation}
    b = \dfrac{\mu_F}{\mu_L} \delta \label{eq:GCT}
\end{equation}
Here, \(\delta\) denotes the local distance between the surface and the fluid-fluid interface, while the viscosity of the lubricant and the main fluid are represented by the terms \(\mu_L\) and $\mu_F$, respectively. 
GCT is, thus, a conceptual precursor to several studies that embed the effect of lubricating fluid layer into boundary conditions \cite{choudhary2015effective, kumar2016permeability, li2017feature, kumar2018liquid}. This involves using a `shape function' for the local slip length exactly similar to the shape of topography \cite{zhou2013effective, asmolov2013flow, nizkaya2013flow}. However, later studies \cite{schonecker2014influence, nizkaya2014gas} showed that the local slip length distribution is not geometrically similar to the corresponding topography shape, and results deviate even for the shallow rib structures.

The gas-cushion theory is improved by the constant shear-stress model at the interface \cite{schonecker2013longitudinal, schonecker2014influence, schonecker2015assessment} for the rectangular ribbed surface.
This theory is valid for any lubricating fluid but assumes constant shear stress in the fluid subdomains of the reference plane, though broadly corroborated by numerical comparisons, lacks mathematical justification from the equations of viscous fluid flow. 
Using this method, solutions are derived for isolated \cite{schonecker2013longitudinal} and periodic \cite{schonecker2014influence, schonecker2015assessment} undulations in topography. This assumption of constant shear stress at the interface is better than the shear-free and gas-cushion theory but not the accurate one.

The exact solution for flow over lubricant-filled ribbed surfaces was also obtained \cite{ng2010effects, sun2017effective} for both longitudinal and transverse flows in a parallel plate channel. 
These studies used the eigen-function expansion method, similar to the approach of Wang et al. \cite{wang2003flow}. However, these analyses apply only to channels with symmetric lubricant-infused surfaces (LIS) on both walls, a configuration that is, on the one hand, unrealistic due to the challenges of maintaining symmetry and alignment at the microscale and, on the other hand, unable to separate the effects of confining the flow passage from the effect of the ribs. Most experimental studies focus on configurations with LIS on just one wall \cite{ou2004laminar, ou2005direct, davies2006laminar, maynes2007laminar, tsai2009quantifying, bolognesi2014evidence, bhandaru2017programmable}. Additionally, knowing the slip length on an isolated ribbed wall can be used to synthesize predictions of permeability in multiple confined configurations (barring situations where the ribs significantly constrict the flow passage). In contrast, a symmetric-channel prediction applies only to a specialized setting.
Hence, the constant shear stress model is the most accurate theory available until now for predicting the slip length for flow over LIS.

In this work, domain decomposition and eigenfunction expansion method were used, and the work of Wang \cite{wang2003flow} and their subsequent work \cite{ng2010effects,ng2011effective,sun2017effective} is extended for shear-driven multiphase flow. 
The analytical solutions are provided for the flow along both the principal directions, \textit{i.e.,} both along and across the ribs. The analytical results are also compared with the numerical simulations and results available in the literature.
Later, the analytical solution is used to analyze the effect of fluid properties and various dimensional parameters, such as rib height and pitch length, on slip. Finally, suggestions are provided for the design of LIS and the choice of lubricating fluid for various applications.

\section{Mathematical model}   \label{sec:MathRec}
\begin{figure}
    \centering
    \begin{subfigure}[t]{0.48\textwidth}
        \centering         
        \includegraphics[width=\textwidth]{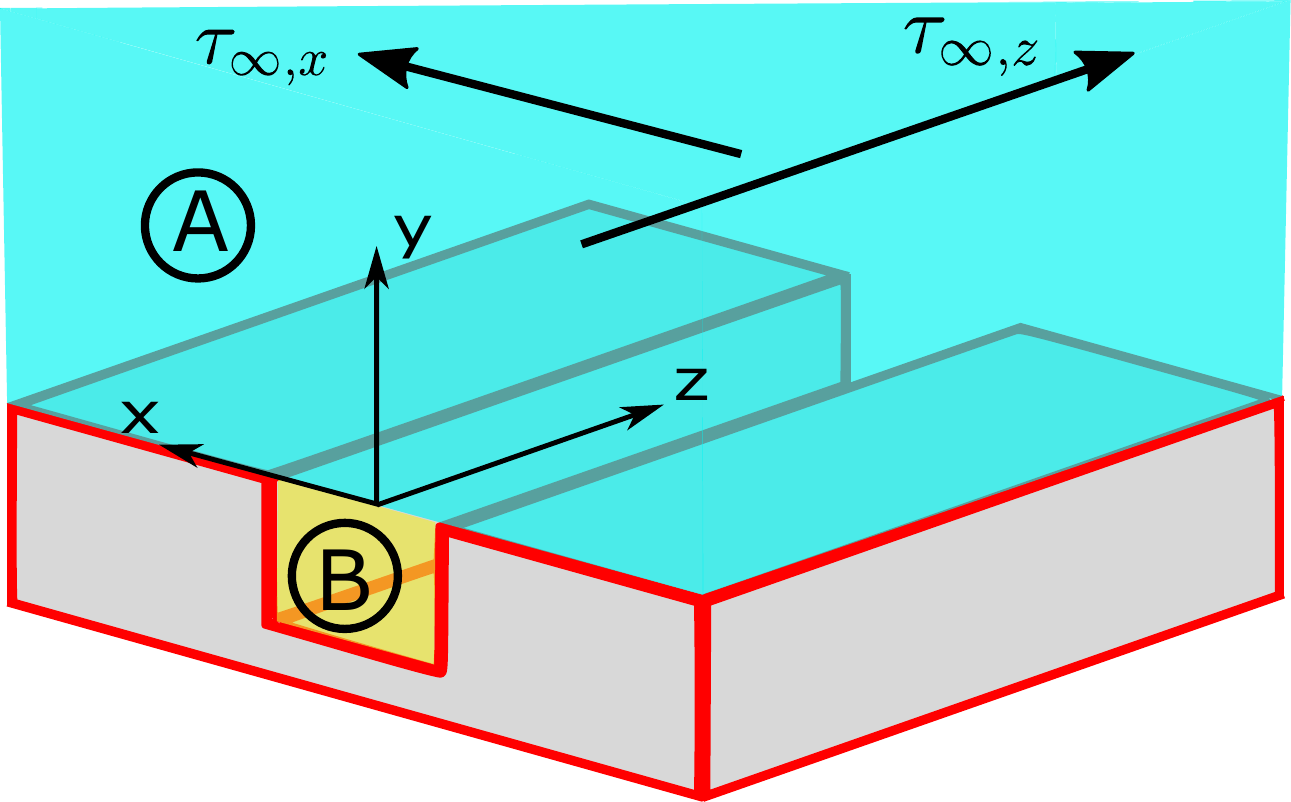}         \caption{ } \label{fig:Schem1SS}     
        \end{subfigure}
    \quad
    \begin{subfigure}[t]{0.48\textwidth}
        \centering         
        \includegraphics[width=\textwidth]{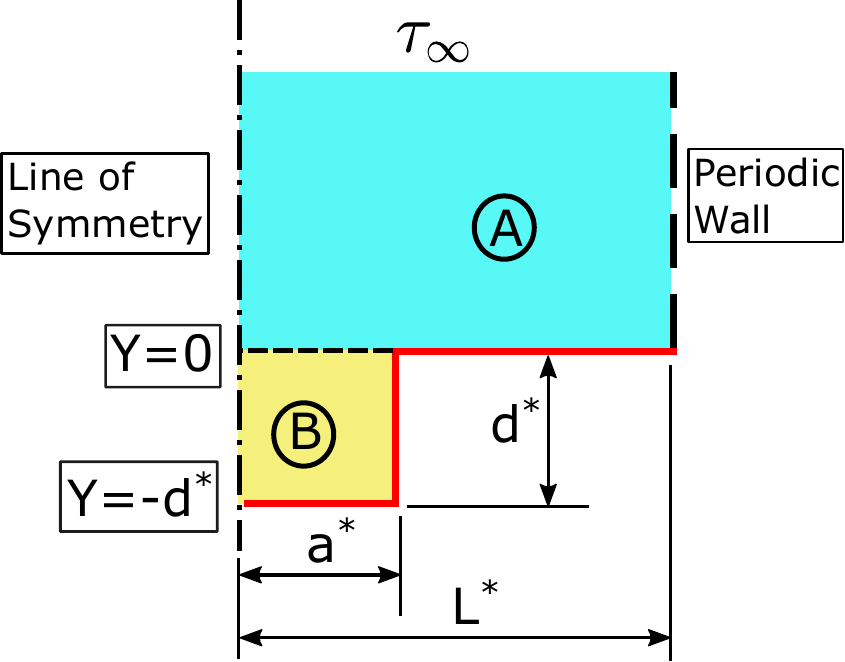}
        \caption{ } \label{fig:Schem2SS}
    \end{subfigure} 
  \caption{Schematic diagram of (a) surface having ribs, with driving force applied along two principal directions, and (b) details of the corrugation geometry. As the corrugations are periodically repeating along $x-$direction, hence only one cell is shown in Figure (a). Due to the symmetric nature of the problem along the mid-plane, figure (b) shows only one-half of the cell along with the boundary condition used to solve the problem.} \label{fig:SchemSS}
\end{figure}

Consider the flow of an incompressible liquid $A$ of viscosity $\mu^A$ over the ribbed surface filled with another immiscible fluid $B$. The flow is assumed to be driven by a constant shear stress ($\tau_\infty$).
The topography is composed of periodically repeating identical rectangular ribs of depth $d^*$, trough width $2a^*$, and pitch length $L$ as shown in Fig. \ref{fig:SchemSS}. The lubricant fluid $B$ of viscosity $\mu^B$ is assumed to be filled exactly up to the peaks of the ribs. The interface between the two fluids is assumed to be flat, neglecting its curvature on the grounds that interfacial curvature has been demonstrated to have a negligible impact on the flow field in similar flow configurations \cite{crowdy2017perturbation, ge2018effective}, and further because the interface becomes planar at steady state for sufficiently large values of the interfacial tension between the two fluids (measured, say, in units of $\tau_\infty/(\mu^A L)$). 

The surface corrugations are anisotropic. Hence, the direction of external force application significantly affects the flow field. However, the analysis is done for the applied force along two principal directions $X$ and $Z$ and the resultant flows are termed as transverse and longitudinal flows, respectively. The flow for the case of driving force applied along any other direction could be obtained with the help of transverse and longitudinal flow using a tensorial formulation \cite{bazant2008tensorial}.


Due to the repeating nature of the topography, it is sufficient to consider a single periodic cell of length $L$, as shown in Fig. \ref{fig:SchemSS}. 
The origin of the axis is selected at the middle of the fluid-fluid interface such that $X=0$ separates the cell in two equal halves, as shown in Fig. \ref{fig:Schem2SS}. 
Because of geometric similarity, the solution is symmetric about the mean line passing through $X=0$, hence, the analytical solution is obtained only for half period cell (from $X=0$ to $+L/2$). For analysis, an eigen-function expansion method, similar to that used by Wang \cite{wang2003flow} and subsequent works \cite{ng2009stokes,ng2010effects,ng2011effective,sun2017effective}, is adopted. For the analysis, the flow domain is divided into two parts by the hypothetical plane passing through the peaks of the ribs, as shown in Fig. \ref{fig:SchemSS}. This plane physically separates the domain of both fluids and is also called a reference plane. 
The domain $A$ is the domain of the main fluid extending in the positive $Y$ region above $Y=0$, and the domain $B$ is the domain of lubricant fluid from $Y=0$ to $-d^*$. 

For steady-state, low Reynolds number flow, the governing Stokes equation for longitudinal flow is given by:
\begin{equation}
    \frac{\partial^2 W^j}{\partial X^2} + \frac{\partial^2 W^j}{\partial Y^2} =0   \label{eq:RectPGovEq}
\end{equation}
Here, the superscript $j=A, B$ is used to differentiate the working fluid $A$ from the lubricant $B$. For transverse flow, the flow and the pressure-gradient field have components along both $X$ and $Y$; the corresponding governing equations can be represented more compactly by the stream-function approach. If $U,V$ denote velocity components along $X,Y$, then $U=\frac{\partial \psi^* }{\partial Y}$ and $V=-\frac{\partial \psi^* }{\partial X}$. With $\nabla^*$ operator defined to have Cartesian components $(\frac{\partial}{\partial X},\frac{\partial}{\partial Y})$, the stream-function $\psi^*$ in each fluid follows
\begin{equation}
    \nabla^{*4} \psi^{*j} = 0            \label{eq:RectTGovEq}
\end{equation}
 The no-slip boundary condition is considered at fluid-solid interfaces, while velocity and shear-stress balance conditions are applied at the fluid-fluid interface. Constant driving shear stress is used as the far-field boundary condition.

Following the conversion of the governing equations and boundary conditions into dimensionless forms, their solution will be obtained using eigenfunction expansions in the following subsections. For non-dimensionalization, the half-pitch length $L/2$ is used as the length scale and $\tau_\infty L/2\mu^A$ as the velocity scale. The effect of far-field shear rate applied parallel to the ribs is investigated under `Longitudinal Flow'' in Sec. \ref{sec:RecShearParallel}.  Section \ref{sec:TransShearFlowRec} (``Transverse Flow'') investigates the case of shear rate applied perpendicular to the ribs.

\subsection{Longitudinal flow} \label{sec:RecShearParallel}
To begin with, we investigated the case of flow driven by shear stress applied along the rib length ($z-$ direction), for which the governing equation is given by Eq. \eqref{eq:RectPGovEq}. 
The non-dimensional form of the corresponding equation is given as:
\begin{equation}
    \frac{\partial^2 w^j}{\partial x^2} + \frac{\partial^2 w^j}{\partial y^2} = 0   \label{eq:RectShearPGovEq}
\end{equation}
It can be noted here that steady shear flow along the rib length does not possess either applied or induced pressure gradient in the flow, and accordingly, the corresponding terms are set to zero in Eq. \eqref{eq:RectShearPGovEq}. On the other hand, pressure gradients are induced in the transverse flow problem, which will be studied later.

The boundary and interfacial conditions on the constant $y$ planes are:
\begin{subequations}
    \begin{align}
        \frac{\partial w^A}{\partial y}\bigg|_{y \rightarrow \infty} & = 1   \label{eq:RecShearInfiP}  \\
        w^A\big|_{y=0} &= w^B\big|_{y=0}   \label{eq:VelBalParallel} \\
        \mu \frac{\partial w^A}{\partial y}\bigg|_{y=0} &= \frac{\partial w^B}{\partial y}\bigg|_{y=0}  \label{eq:StressBalParallel} \\
        w^B|_{y=-d} &= 0 .   \label{eq:RecShearWallP}
    \end{align}  \label{eq:RecShearPBC}
\end{subequations}
Here, $\mu$ is the viscosity ratio ($\mu = \mu^A/\mu^B$), which can also be interpreted as the viscosity of the working fluid is normalized by that of the lubricant. Thus, $\mu < 1$ or $>1$, according as the lubricant is, respectively, more viscous or less viscous than the working fluid.
The first boundary condition represents the driving, constant shear stress applied far from the wall. The second and third equations represent the continuity of velocity and shear stress at the fluid-fluid interface. The final equation represents the no-slip condition for the lubricant at the bottom wall of the topography.

The velocity distribution in the two fluids possesses the following eigenfunction expansions:
\begin{subequations}
    \begin{align}
        w^A &= y + w_s + \sum_{n=1}^{n=\infty} { \cos{(\alpha_n x)}  A_n e^{-y\alpha_n}   } \label{eq:aVelExpShearPRec}  \\
        w^B &= \sum_{n=1}^{n=\infty} { \cos{(\beta_n x)} B_n \left(  e^{y\beta_n} -  e^{- \beta_n (y+ 2d)}  \right)  }.
  \label{eq:bVelExpShearPRec}  \end{align}  \label{eq:VelExpShearPRec}
\end{subequations}
Here, the variables $\alpha_n=n\pi$ and $\beta_n = \frac{(2n-1)\pi}{2a}$, and the summation index $n$ varies from $+1$ to $+\infty$ in this chapter, and accordingly $\displaystyle\sum_{n=1}^\infty$ would be given the abbreviated notation $\displaystyle\sum_n$. 
The first term on the right-hand side of Eq.
\eqref{eq:aVelExpShearPRec} 
represents the velocity component in the absence of ribs.
The second term is the average effect of the ribs and lubricating fluid on the composite fluid-solid plane $y=0$. This is the net average velocity at $y=0$, termed the slip velocity. The last term is the local variation due to the ribs and lubricant. The average of the last term over the surface is zero. The velocity of lubricant, calculated from \eqref{eq:bVelExpShearPRec}, consists of only terms that arise from the topography variation. 
The above functional forms of the velocity profiles automatically satisfy the far-field and wall boundary conditions (Eq. \eqref{eq:RecShearInfiP} and \eqref{eq:RecShearWallP}). The other two conditions at the interface need to be applied to obtain the constants $A_n$ and $B_n$.

Multiplying the velocity balance condition \eqref{eq:VelBalParallel} with $\cos{(\alpha_m x)}$ by substituting the expressions from Eq. \eqref{eq:VelExpShearPRec} and integrating from $x=0$ to $1$, we get: 
\begin{align}
    \frac{A_m}{2} = \sum_n { I_{mn} B_n \left(  1 -  e^{- 2d \beta_n  }  \right)  }      \label{eq:ShearVelBal}
\end{align}
here the parameter $I_{mn}$ is given by
\begin{align}
    I_{mn} =
 \left\{
     \begin{aligned}
       \frac{\sin (a (\alpha_m -\beta_n ))}{2 (\alpha_m -\beta_n )} &+\frac{\sin (a (\alpha_m +\beta_n ))}{2 (\alpha_m +\beta_n )} \hspace{40 pt} &\text{for $\alpha_m \neq \beta_n$} \\
      & \frac{a}{2} \hspace{50 pt} &\text{for $\alpha_m = \beta_n$}
     \end{aligned}
     \right.   \label{eq:ImnValue}
\end{align}

Next multiplying the stress-balance equation \eqref{eq:StressBalParallel} with $\cos{(\beta_m x)}$ and integrating from $x=0$ to $a$ give:
\begin{align}
    B_m \frac{a \beta_m}{2} \left(  1 +  e^{- 2d \beta_m }  \right) = \frac{ (-1)^{m+1} \mu }{\beta_m} - \sum_n { I_{nm} \alpha_n   A_n  }  \label{eq:ShearStressBal}
\end{align}
where the term $I_{nm} = I_{mn}$ as given in Eq. \eqref{eq:ImnValue}, just by replacing the subscript $m$ with $n$ and vice-versa.

We truncate the summation index of $w^A$ to $M$ terms, and of $w^B$ to $N$ terms, where $N$ is the nearest integer of $N=aM$.
Now, Eq. \ref{eq:ShearVelBal} and \ref{eq:ShearStressBal} together create a system of linear equations with $M+N$ unknowns and equations.
For obtaining the unknowns $A_n$ and $B_n$, this system of linear equations is solved.

After obtaining all $A_n$ and $B_n$'s, the only unknown remains in the slip velocity $w_s$. 
Integrating the velocity balance condition (\ref{eq:VelBalParallel}) from $x=0$ to $1$ give slip velocity as: 
\begin{equation}
    w_s = \sum_n { (-1)^{n+1} \frac{B_n }{\beta_n} \left(  1 -  e^{- 2d \beta_n  }  \right)  }         \label{eq:Wslip}
\end{equation}

This is the semi-analytical expression for slip velocity, which can be obtained after knowing $B_n$ from the previous step. This is the average non-zero velocity at the plane touching the peaks of ribs. The accuracy of $w_s$ depends on the number of terms in the summation. 
After obtaining the slip velocity for shear stress applied along the $z-$ direction, next, we obtain the slip velocity for the case of stress applied along the $x-$ direction.

\subsection{Transverse flow} \label{sec:TransShearFlowRec}
Now, consider the shear stress $\tau_\infty$ applied along the $x-$ direction. With the same scaling parameters as used for longitudinal flow, the non-dimensional governing equation for stream-function is:
\begin{align}
    \nabla^4 \psi^j = 0         \label{eq:GovEqTransRec}
\end{align}

The assumed form of velocity for transverse flow can be given as:
\begin{subequations}
    \begin{align}
        u^A &=   y+ u_s +  \sum_n { \cos{(\alpha_n x)} E_n  e^{-\alpha_n y} \left( 1 - \alpha_n y   \right)}   \label{eq:uVelShearTRec}    \\
        v^A &=  \sum_n { \frac{\sin{(\alpha_n x)}}{\alpha_n} E_ny e^{-\alpha_n y}  }             \\
        u^B &= \sum_n {\cos{(\beta_n x)} \big( B_{1n}  F_n'(y) + B_{2n} G_n' (y)  \big) } + \sum_n {D_n H_n(x) \gamma_n \cos{(\gamma_n y)}  }    \label{eq:VelExpShearTRecBu}  \\
        v^B &= \sum_n {\sin{(\beta_n x)} \beta_n \bigg( B_{1n} F_n(y) + B_{2n} G_n (y)  \bigg) } - \sum_n {D_n H_n^{'}(x) \sin{(\gamma_n y)}  }.     \label{eq:VelExpShearTRecBv}
    \end{align} \label{eq:VelExpShearTRec}
\end{subequations}
In the above equations, $\alpha_n$ and $\beta_n$ are the same as used for longitudinal flow and $\gamma_n = \frac{n \pi}{d}$, and the superscript $'$ indicates differentiation of a function with respect to its independent variable. The coefficients $B_{1n}$, $B_{2n}$, $D_n$, and $E_n$ are unknown, which will be obtained as part of the solution, while the functions $F_n$, $G_n$ and $H_n$ are defined as follows:
\begin{subequations}
    \begin{align}
        F_n (y) &= \sinh{(\beta_n y)} -\frac{y}{d} \sinh{(d\beta_n)}  e^{-\beta_n (y+d)}         \\
        G_n (y) &= y \sinh{(\beta_n (y+d))}    \\
        H_n(x) &= \left(e^{\gamma_n(x-a)} +e^{-\gamma_n (x+a)} \right)- \frac{x (1+e^{-2 a \gamma_n }) (e^{\gamma_n (x-a)} -e^{-\gamma_n(x+a)})}{a(1-e^{-2 a \gamma_n })}. 
    \end{align}
\end{subequations}

The velocity must satisfy the following non-dimensional boundary conditions:
\begin{subequations}
    \begin{align}
        \frac{\partial u^A}{\partial y}\bigg|_{y \rightarrow \infty} &= 1     \label{eq:ShearStressS} \\
        u^A|_{y=0} &= u^B|_{y=0}     \label{eq:VelBalTransS}  \\
        \mu \frac{\partial u^A}{\partial y}\bigg|_{y=0}  &= \frac{\partial u^B}{\partial y}\bigg|_{y=0}       \label{eq:StressBalTransS}    \\
        u^B|_{y=-d} &= 0       \label{eq:GrooveBottomNoSlipS}      \\
        v^B|_{x=a}  &=0         \label{eq:GrooveSideNoSlipS}
    \end{align}
\end{subequations}
The first four conditions are the same as used for the longitudinal flow out of which the first condition Eq. \eqref{eq:ShearStressS} of constant shear-stress is satisfied by the main fluid velocity profile. The fifth condition is the additional, which is not present in the case of longitudinal flow. This condition satisfies the no-slip for the $v$-velocity component of lubricant on the rib's vertical wall. For longitudinal flow, the lubricant's velocity component along $y-$direction is absent. Therefore, this condition does not appear for longitudinal flow. We will now apply the boundary conditions \eqref{eq:VelBalTransS} to \eqref{eq:GrooveSideNoSlipS} to obtain the unknown constants in the velocity profiles.

First, multiplying the no-slip condition on the vertical wall of the ribs (Eq. \eqref{eq:GrooveSideNoSlipS}) with $\sin{(\gamma_m y)}$ and integrating from $y=-d$ to $0$, the following is obtained:
\begin{align}
    0 = \sum_n  2 &\beta_n^2 \gamma_m  (-1)^{n+1} \bigg( B_{1n}  \frac{   \tanh{(d \beta_n )} \left(e^{-\beta_n  d} - (-1)^{m}\right)}{d \left(\beta_n ^2+\gamma_m ^2\right)^2}  \nonumber   \\
    &+ \!\! B_{2n} \frac{   ( 1- sech(d \beta_n ) (-1)^{m})}{\left(\beta_n ^2+\gamma_m ^2\right)^2}   \bigg)      + D_m \frac{d}{2} \!\! \left(\frac{1+4 a \gamma_me^{-2 a \gamma_m } -e^{-4 a \gamma_m } }{a (1- e^{-2 a \gamma_m })} \right)    \label{eq:ShearTransDn}
\end{align}

Next, multiplying the no-slip condition at the bottom surface ($y=-d$) \eqref{eq:GrooveBottomNoSlipS} with $\cos{(\beta_m x)}$ and integrating from $x=0$ to $a$
\begin{align}
    0  =  \frac{a \beta_m}{2} &\bigg( B_{1m}   \bigg( 1 -  \tanh{(d \beta_m)} \left(1+ \frac{1}{d \beta_m } \right)\! \bigg) - B_{2m} d \: sech(d \beta_m) \bigg) + \sum_n {  \gamma_n (-1)^{n} D_n H_{2n}    }  \label{eq:ShearTransB2n}
\end{align}
where the coefficient $H_{2n}$ is given as
\begin{subequations}
    \begin{align}
        H_{2n}&= \int_0^a \cos{(\beta_m x)} H_n(x)    \\
        &= \frac{ 2\beta_m  \gamma_n   (-1)^{m+1} \left(  1 + e^{-2a \gamma_n } \right)^2  }{ a \left(\beta_m ^2+\gamma_n ^2\right)^2  \left(  1 - e^{-2a \gamma_n } \right) }
    \end{align}   \label{eq:H2nValue}
\end{subequations}

Further, multiplying velocity balance condition at interface \eqref{eq:VelBalTransS} with $\cos{(\alpha_n x)}$ and integrating from $x=0$ to $1$, we get
\begin{align}
    \frac{E_m}{2} = \sum_n {D_n H_{3mn} \gamma_n  } +\sum_n  I_{mn} \bigg(  B_{1n} \beta_n & \left( sech{(d \beta_n)} -\frac{e^{-d \beta_n } \tanh{(d \beta_n)} }{d \beta_n } \right) \nonumber \\
    & + B_{2n}    \tanh{(d \beta_n)}  \bigg)    \label{eq:ShearTransEn}
\end{align}
here, $I_{mn}$ is same as given by Eq. \eqref{eq:ImnValue}, and the term $H_{3nm} $ is given by:
\begin{subequations}
    \begin{align}
        H_{3nm} =& \int_0^a { H_n \cos{(\alpha_m x)} dx   }   \\
        =& \frac{  \cos (a \alpha_m ) \big((\gamma_n^2 -\alpha_m^2 )  (1-e^{-4 a \gamma_n })-4 a \gamma_n e^{-2 a \gamma_n } \left(\gamma_n ^2+\alpha_m ^2\right)\big)}{ a (1 -e^{-2a \gamma_n} )  \left(\gamma_n ^2+\alpha_m ^2\right)^2}   \nonumber \\
        &\; \; \;\; \; \;\; + \frac{ 2 \gamma_n  \alpha_m  (1 +e^{-2a \gamma_n} )^2 \sin (a \alpha_m )}{a (1 -e^{-2a \gamma_n} )  \left(\gamma_n ^2+\alpha_m ^2\right)^2}
    \end{align}   \label{eq:H3nmTransflow}
\end{subequations}
where, the term $I_{nm}$ is again can be obtained from Eq. \eqref{eq:ImnValue} by substituting all subscripts $m$ with $n$ and vice versa.

Next, multiplying shear-stress balance equation at interface \eqref{eq:StressBalTransS} with $\cos{(\beta_m x)}$ and integrating from $x=0$ to $a$ we get:
\begin{align}
    a\bigg( B_{1m} \frac{ \beta_m  e^{-d \beta_m  } \tanh {(\beta_m  d)}}{d} + B_{2m}  \beta_m   \bigg) = \frac{\mu (-1)^{m+1} }{\beta_m} -\sum_n  { I_{nm}  E_n 2\mu \alpha_n    }   \label{eq:ShearTransB1n}
\end{align}

Now, similar to longitudinal flow, all summation indexes are truncated to finite terms, which will have a direct influence on the accuracy of the solution. The summation index of $u^A$ is truncated to $M$ terms, and for lubricant, the coefficients $B_{1n}$ and $B_{2n}$ are truncated to $N$ terms, and the summation index for coefficient $D_n$ truncated to $P$ terms. Here, $N$ and $P$ are the nearest integer of $N=aM$ and $P=bM$, respectively. 
Now, we have equal numbers of linear algebraic equations \eqref{eq:ShearTransDn}, \eqref{eq:ShearTransB2n}, \eqref{eq:ShearTransEn}, and \eqref{eq:ShearTransB1n} for same numbers of unknowns. 

After obtaining all unknowns from the system of linear algebraic equations, the slip velocity could be obtained by integrating velocity balance condition at interface \eqref{eq:VelBalTransS} from $x=0$ to $1$, as follows:
\begin{multline}
    u_s = \sum_n {  \frac{(-1)^{n+1}}{\beta_n} \bigg( B_{1n} \beta_n  \left( sech{(d \beta_n)} -\frac{e^{-d \beta_n } \tanh{(d \beta_n)} }{d \beta_n } \right)  + B_{2n}    \tanh{(d \beta_n)}  \bigg) }     \\
    + \sum_n {D_n  \frac{ 1 - e^{-4 a \gamma_n} -4 a \gamma_n e^{- 2a \gamma_n} }{a \gamma_n  (1-e^{-2 a \gamma_n})} }      \label{eq:Uslip}
\end{multline}
This is the semi-analytical expression for slip velocity, with driving force applied across the ribs. With this, we obtain the solution of flow in the thick channel for flow along both principal directions. 
Complete velocity profiles for main and lubricating fluid could be obtained from Eqs. \eqref{eq:VelExpShearPRec} and \eqref{eq:VelExpShearTRec} for longitudinal and transverse flow, respectively.

\section{Results and discussion}\label{sec:RecResult} 
In the previous section, an analytical solution of the locally resolved flow field is obtained using the eigen-function expansion method. The solutions are provided for the flow field of the main fluid as well as for the lubricant filled in the gap between the ribs of the surface, with no specific assumptions made regarding the fluids' properties or rib dimensions.
Using the analytical solution derived in Sec. \ref{sec:MathRec}, the slip length can be obtained for any surface having ribs.
For obtaining apparent slip length over a surface having ribs, the Navier slip length formula given by Eq. \eqref{eq:slipLength} is modified by Bazant and Vinogradova \cite{bazant2008tensorial} and given as:
\begin{equation}
    b_\parallel \big|_{y=0} = \frac{\langle w^A \rangle }{\langle \partial w/\partial y \rangle }\bigg|_{y=0};  \hspace{20 pt} \textit{and} \hspace{20 pt} b_\perp \big|_{y=0} = \frac{\langle u^A \rangle }{ \langle \partial u/\partial y \rangle }\bigg|_{y=0}    \label{eq:SlipLengthFormula}
\end{equation}
Here, terms inside the parentheses \( \langle \dots \rangle \) are the spacial average of any quantity \( \dots \) over the pitch length. As the ribs are identical and repeating, an average over the pitch is equal to the surface average.
Also, the subscripts \(\parallel\) and \( \perp \) indicate the surface average slip length for flow along and across the ribs.

Surface average velocity can be obtained by eliminating variable terms in Eqs. \eqref{eq:aVelExpShearPRec} and \eqref{eq:uVelShearTRec} for longitudinal and transverse flow, respectively. Differentiating the same equations will provide the velocity gradient for respective flows. The numerator term at peaks of ribs is equal to the surface average slip velocity, while the surface average denominator term will be unity. Hence, slip length for shear-driven flows is dependent only on the slip velocity.

In the following sections, first, the local variation of velocity at the reference plane, obtained from the eigenfunction expansion-based theory, will be subjected to stringent comparisons with alternative numerical approaches. After validating the accuracy of the analytical solution, the effect of various fluid and surface parameters on slip length will be analyzed.

\subsection{Numerical comparisons\label{sec:schoe}}
As the slip length depends only on the fluid velocity at the reference plane, first in this section, we will compare the accuracy of the analytical solution by comparing it with the numerical results. 
For longitudinal flows, the analytical solutions are validated against numerical solutions obtained using COMSOL Multiphysics software. While, results presented by Sch\"{o}necker et al. \cite{schonecker2014influence} are used to check the accuracy of the analytical results for transverse flow. Results are plotted for the cases of viscosity ratio (\( \mu = \mu_A/\mu_B\)) of 50 and 0.02. 
The first case of $\mu=50$ shows the case of water as the main fluid and lubricating fluid as air, representing superhydrophobic surfaces (SHS). The other case ($\mu=0.02$) represents the opposite scenario, with water as the lubricating fluid and air as the main fluid. Most practical combinations of fluids come under these extreme $\mu$ ranges.

Longitudinal flows (flow along $z-$ direction) on ribbed surfaces are simulated on COMSOL using the coefficient-form PDE module of the software. This module is used to solve the governing Eq. \eqref{eq:RectShearPGovEq}.
Simulations are conducted for the entire cell of pitch length $L$, and the periodic boundary condition is applied at the ends of the cell, relaxing the symmetry boundary condition at the center of the domain. The no-slip condition is applied at all solid-fluid interfaces, and at fluid-fluid interfaces, the velocity and stress balance conditions are applied. 

\begin{figure}[htbp]
    \centering
    \begin{subfigure}[t]{0.48\textwidth}
        \centering         
        \includegraphics[width=\textwidth]{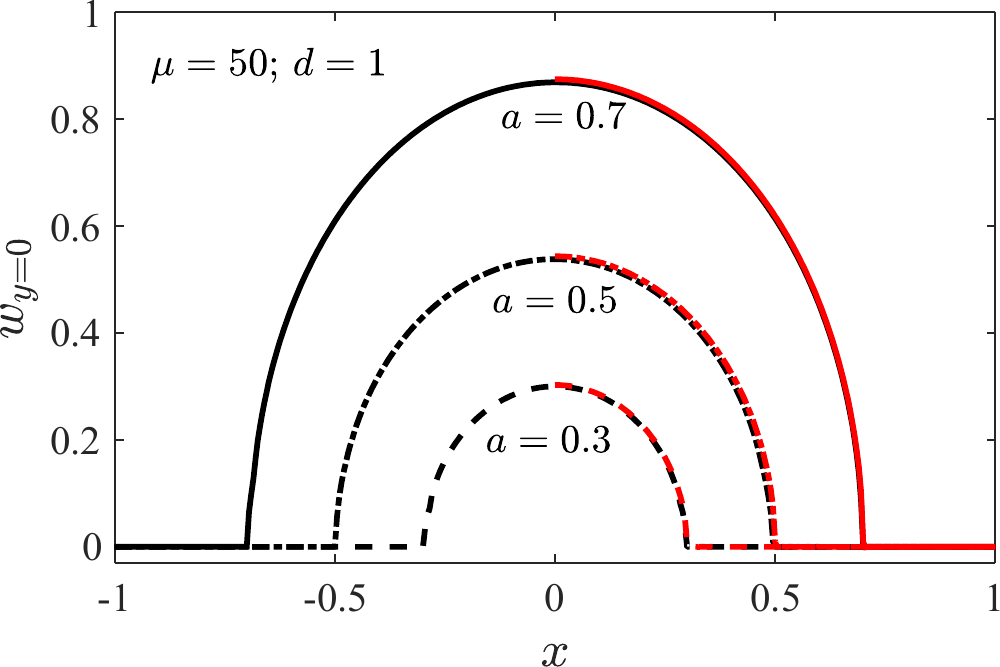}         \caption{ } 
        \end{subfigure}
    \quad
    \begin{subfigure}[t]{0.48\textwidth}
        \centering         
        \includegraphics[width=\textwidth]{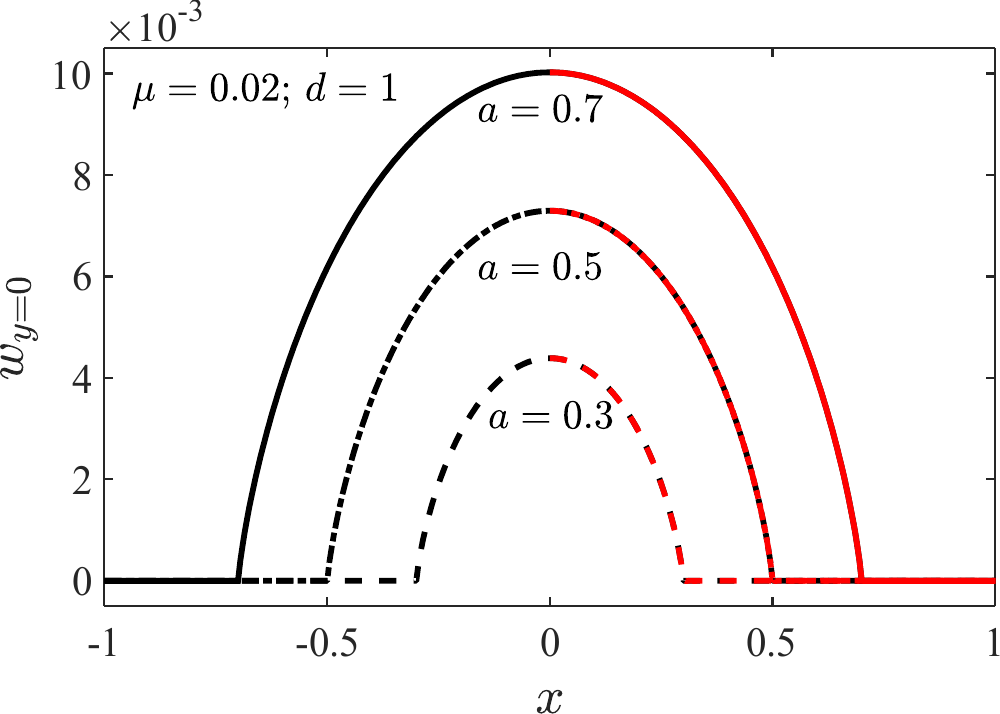}
        \caption{ } 
    \end{subfigure} 
  \caption{Comparison between analytical and numerical results for shear-driven longitudinal flow on rectangular shaped LIS for the case of (a) \(\mu = 50\), and (b) \(\mu=0.02\).} \label{fig:LocalVelShear}
\end{figure}

The comparison between analytically and numerically obtained velocity at the reference plane for longitudinal flows is shown in Fig. \ref{fig:LocalVelShear}. Results are plotted for various lubricant fractions ($a$), keeping the height of ribs ($d=1$) same for all cases. 
The analytical solution is plotted only for half a cell, while the numerical results are plotted for the full cell length. Due to the symmetry of the velocity profile about mid-plane, the velocity for the other half of the cell is exactly the same. 
Both analytical and numerical solutions, when plotted for the full period length, overlap with each other. Therefore, for better visualization, the analytical result is plotted for only half the cell length.
Black lines show the numerical results and analytical velocity is plotted in red color. Fig. \ref{fig:LocalVelShear} shows that the analytical model accurately predicts the velocity at the reference plane for all considered cases.

\begin{figure}
    \centering
    \begin{subfigure}[t]{0.48\textwidth}
        \centering         
        \includegraphics[width=\textwidth]{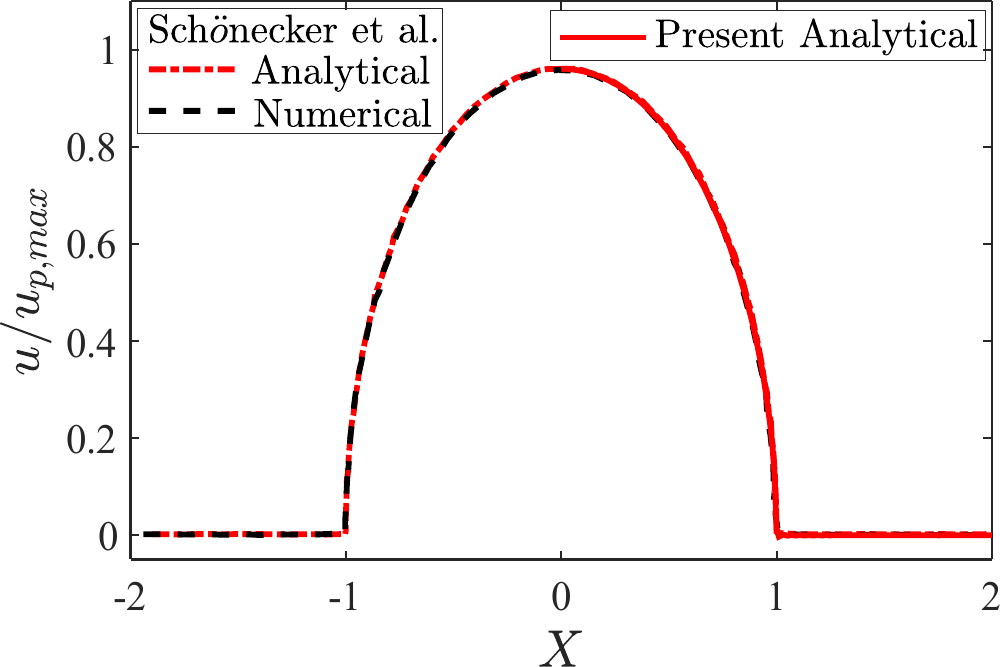}         \caption{ } 
        \end{subfigure}
    \quad
    \begin{subfigure}[t]{0.48\textwidth}
        \centering         
        \includegraphics[width=\textwidth]{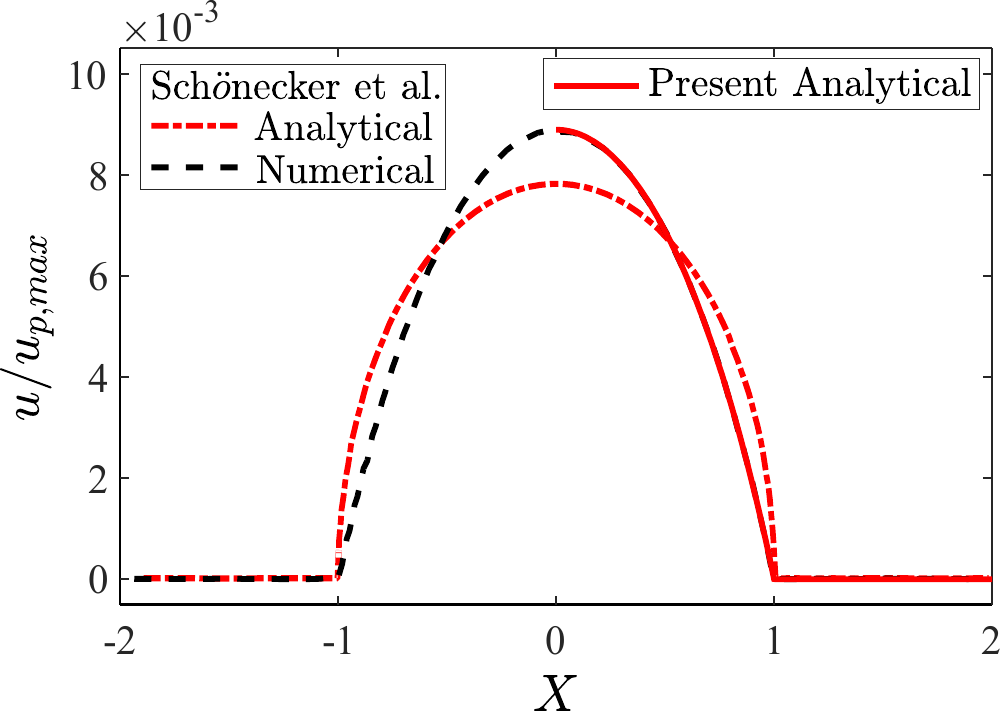}
        \caption{ } 
    \end{subfigure} 
  \caption{Comparison between analytical and numerical results for transverse flow on rectangular shaped LIS for the case of (a) \(\mu=55\) and (b) \(\mu=0.02\).} \label{fig:SchoneckerCompare}
\end{figure}

Velocity at the reference plane for transverse flow is shown in Fig. \ref{fig:SchoneckerCompare}.
The present analytical results for transverse flow are compared with the analytical and numerical results of Sch\"{o}necker et al. \cite{schonecker2014influence}. Results are plotted for normalized transverse velocity at the reference plane for both cases of $\mu=55$ and $0.02$. The transverse velocity is normalized with the velocity $u_{p,max}=(Im ( \arccos(\sec(a\pi/2)))/2\pi)$, which is the velocity at reference plane given by Philip \cite{philip1972flows, philip1972integral} with shear-free interface assumption.

The analytical model of Sch\"{o}necker et al. \cite{schonecker2014influence} holds the assumption of constant shear stress at the fluid-fluid interface, whereas their numerical work is free from such assumptions. Therefore, the numerical results are considered as a reference for the current analysis. 
The comparison of current transverse flow results from the numerical work of \cite{schonecker2014influence} shows an exact match for slip velocity, while their own analytical model deviates for the case of $\mu=0.02$, because of constant shear-stress assumption. 

Thus, the comparison of the current analytical model with numerical and literature results confirms its ability to precisely predict the velocities of both the main and lubricant fluids across various fluid choices and rib dimensions. 
Having validated our analytical model, we proceed to investigate the surface's slip characteristics under various parameter variations. Finally, we explore the anisotropic slip properties of the surface, comparing the slip along both principal directions.

\subsection{Slip length}   \label{sec:RecSlipLength}
Velocity prediction in Sec. \ref{sec:MathRec} can be used to predict the slip length using the formula \eqref{eq:SlipLengthFormula} for various geometric and fluid parameters. Geometric parameters include the height of the ribs and the gap between them, while the fluid parameters include the viscosity ratio (\(\mu\)) between both fluids.
It is evident from the Figs. \ref{fig:LocalVelShear} and \ref{fig:SchoneckerCompare}, that for both longitudinal and transverse flow, the slip velocity is much larger for low-viscosity lubricants compared to higher ones. Hence, it can be concluded that the viscosity of lubricant negatively affects the slip velocity and slip length.

For a rectangular rib with a specified pitch, the lubricant fraction $a$ (which is also the gap between two ribs) and the rib height ($d$) are the two key parameters that can be optimized to maximize the slip length. 
Both these dimensions directly influence the volume of lubricant trapped between ribs positively.
Hence, it is intuitive to expect that increasing either parameter would lead to an increase in the slip length, which is not only borne out by the current theoretical findings (not shown), but the literature on related problems \cite{philip1972flows,schonecker2014influence,nizkaya2014gas}. Here, we explore a relatively more intriguing yet practically motivated situation,  where $a$ and $d$ are varied, holding the volume of the lubricant $v=2da$ fixed.  For a constant volume of lubricant, increasing lubricant fraction ($a$) leads to a decrease in the rib height ($d$).

\begin{figure}[htbp]
    \centering
    \begin{subfigure}[t]{0.48\textwidth}
        \centering         
        \includegraphics[width=\textwidth]{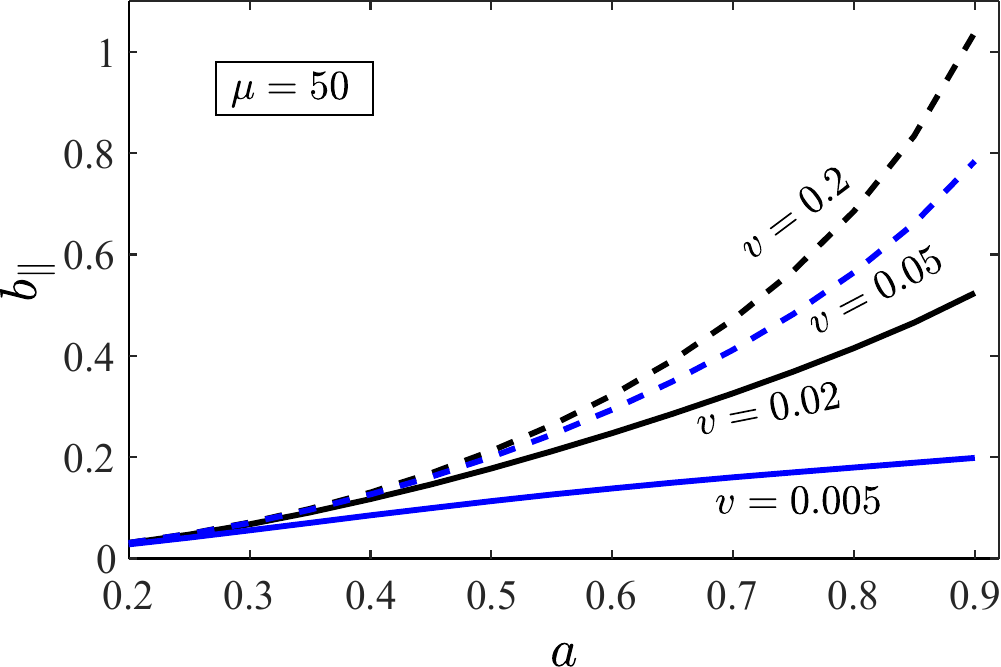}         \caption{ } \label{fig:VolRecPMu50}     
        \end{subfigure}
    \quad
    \begin{subfigure}[t]{0.48\textwidth}
        \centering         
        \includegraphics[width=\textwidth]{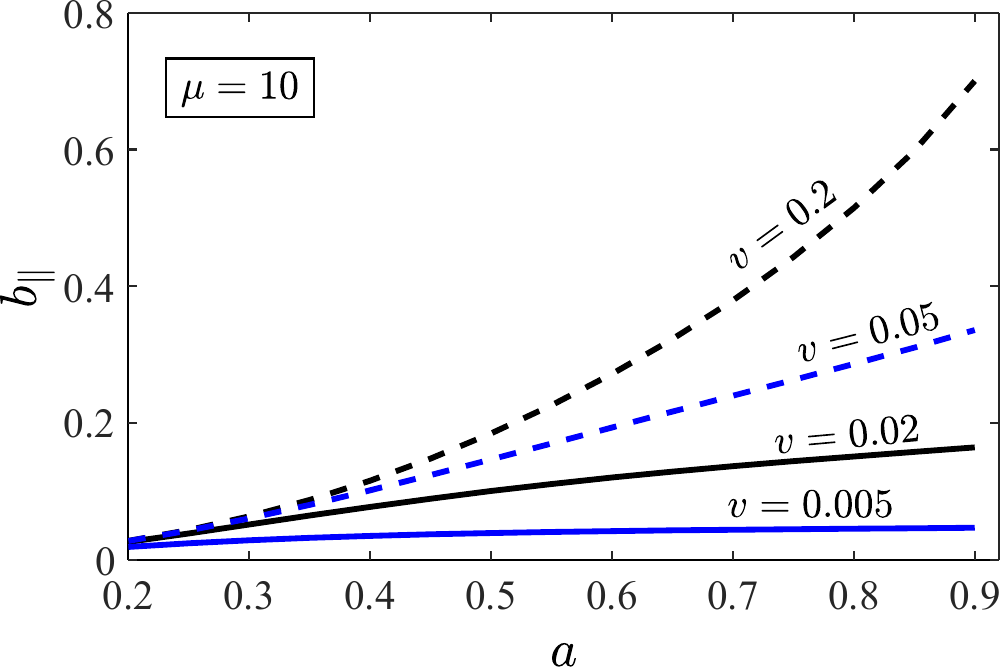}
        \caption{ } \label{fig:VolRecPMu10}
    \end{subfigure} 
    \begin{subfigure}[t]{0.48\textwidth}
        \centering         
        \includegraphics[width=\textwidth]{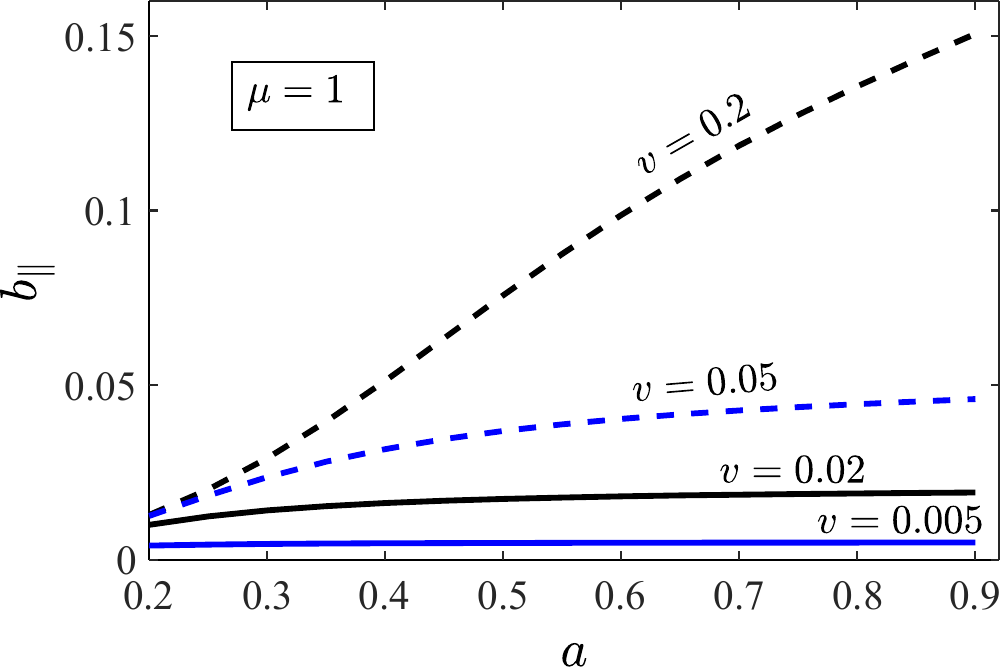}         \caption{ } \label{fig:VolRecPMu01}     
        \end{subfigure}
    \quad
    \begin{subfigure}[t]{0.48\textwidth}
        \centering         
        \includegraphics[width=\textwidth]{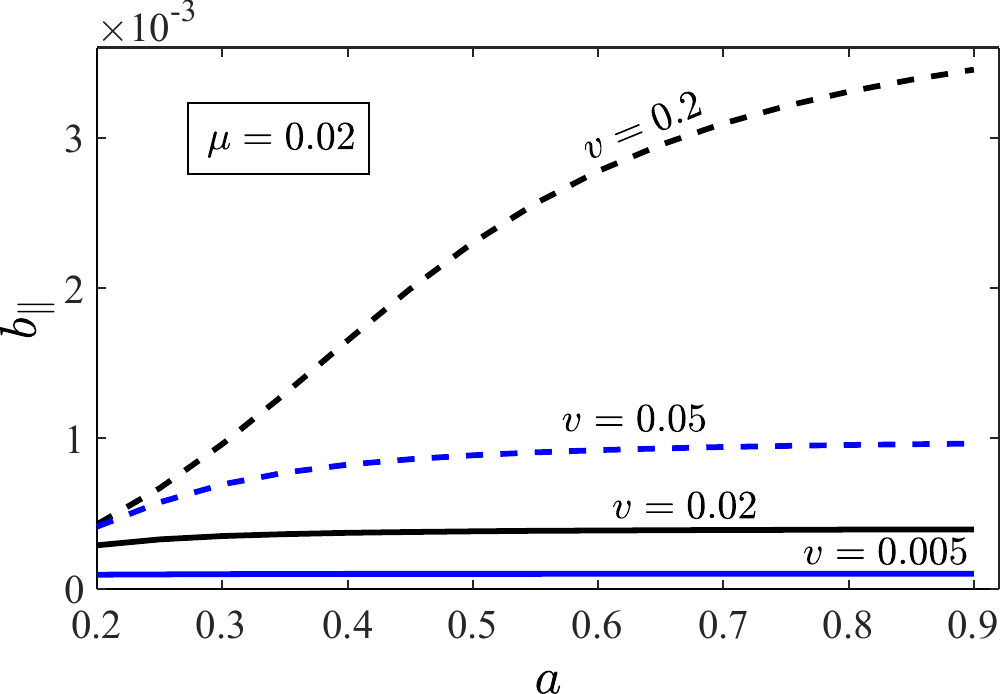}
        \caption{ } \label{fig:VolRecPMu0_02}
    \end{subfigure} 
  \caption{Variation of slip length for the flow along $z-$ direction with the lubricant fraction ($a$) at the constant lubricant volume ($v=2da$) for the case of (a) $\mu=50$, (b) $\mu=10$, (c) $\mu=1$, and (d) $\mu=0.02$.} \label{fig:ParallelVolEffect}
\end{figure}

For longitudinal flow, the results are plotted for the four different values of the viscosity ratio, $\mu=50$, $10$, $1$, and $0.02$. It can be observed that for the high-viscosity lubricant ($\mu = 1$ and $0.02$), the slip length is nearly constant for small lubricant volume ($v=0.005$, $0.02$, and $0.05$) and is not affected by the lubricant fraction (or rib depth). However, at large lubricant volume ($v=0.2$) of the high viscosity lubricant, slip length increases rapidly with the increase in the lubricant fraction, even if the rib height ($d$) decreases with an increase in lubricant fraction ($a$).  
For the lower viscosity lubricant, the trend of increasingly faster rise in slip length with an increase in lubricant fraction (notwithstanding the accompanying decrease in $d$)  holds true even for low lubricant volume ($v$), as shown in Fig. \ref{fig:VolRecPMu50} and \ref{fig:VolRecPMu10}. Hence, it can be concluded that the lubricant fraction is a more significant factor than the rib height ($d$). Even though the rib height reduces with an increase in the lubricant fraction, the effective slip length either increases with an increase in the lubricant fraction ($a$), or at least remains constant. A decrease in rib height does not affect slip length significantly, considering that $a$ and $d>0$.

\begin{figure}[htbp]
    \centering
    \begin{subfigure}[t]{0.48\textwidth}
        \centering         
        \includegraphics[width=\textwidth]{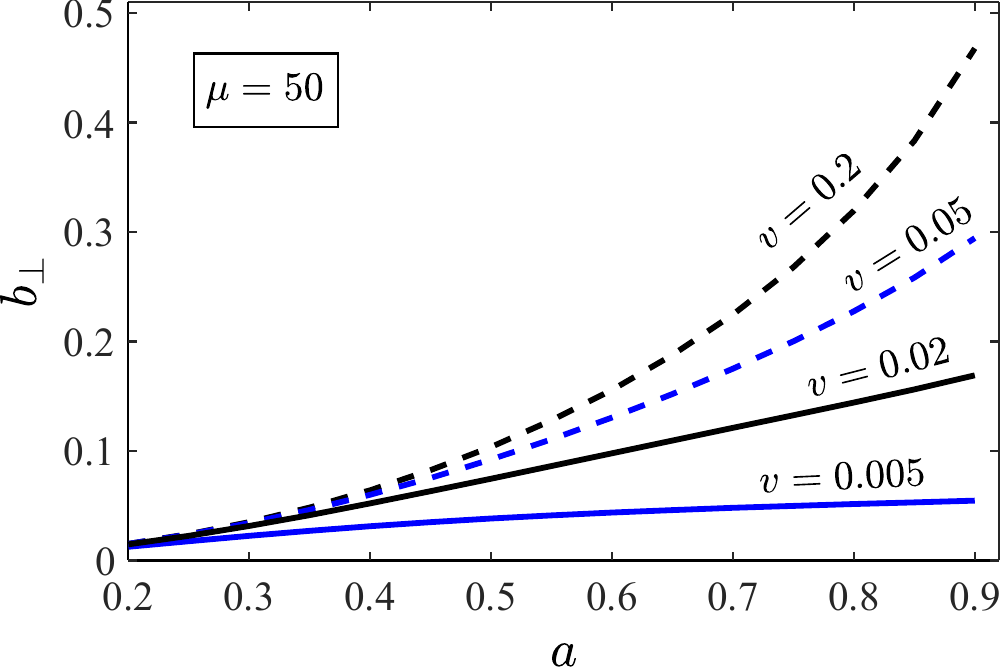}         \caption{ } 
        \end{subfigure}
    \quad
    \begin{subfigure}[t]{0.48\textwidth}
        \centering         
        \includegraphics[width=\textwidth]{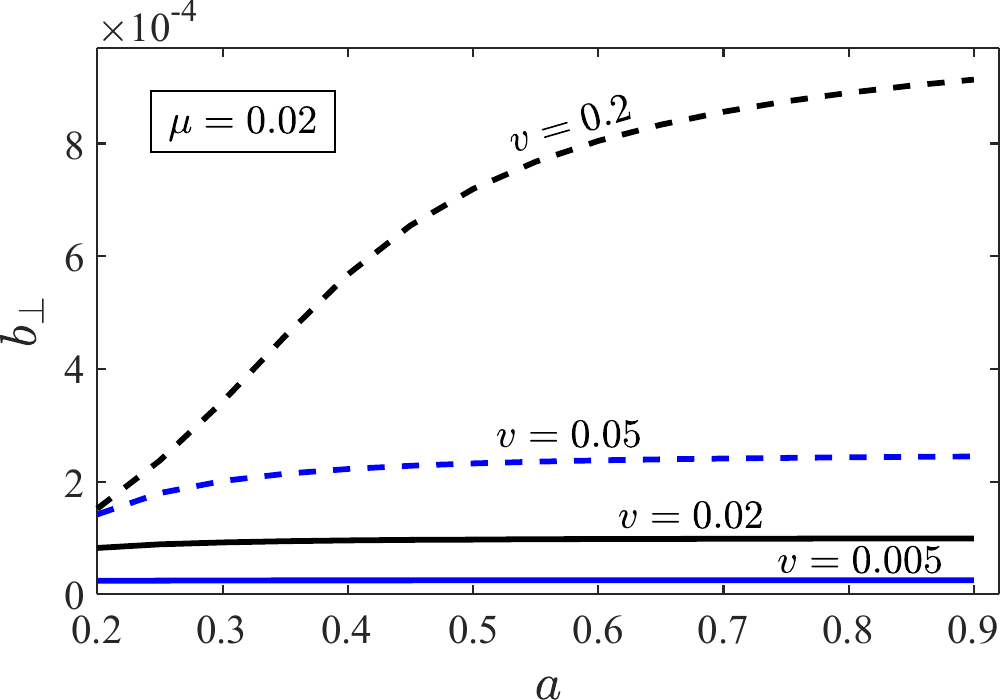}
        \caption{ } 
    \end{subfigure}  
  \caption{Variation of effective slip length for the flow along $x-$ direction with the lubricant fraction ($a$) at the constant lubricant volume ($v=2da$) for the case of (a) $\mu=50$, and (b) $\mu=0.02$.} \label{fig:TransVolEffect}
\end{figure}

Variation of the effective slip length for flow transverse to the ribs at constant lubricant volume is shown in Fig. \ref{fig:TransVolEffect}. Here, the results are plotted only for the two extreme cases of $\mu=50$ and $0.02$. The variation of slip length in transverse flow is qualitatively similar to that in longitudinal flow. The lower viscosity lubricant ($\mu=50$) shows slip length, which grows at increasingly faster rates with increasing lubricant fraction, even though $d$ decreases. However, it takes larger lubricant volumes to get a similar effect with high-viscosity lubricants. The lower the lubricant volume, the smaller the slip, and the quicker the saturation in $a-$ dependence of slip.

Hence, among rectangular-rib patterns chosen to imbibe the same volume of lubricant (possibly limited by cost considerations), the focus for reducing friction should be on replacing fluid-solid contact rather than increasing the depth of the lubricant layer. Smaller values of $d$ can, at worst, lead to saturation of the effect of $a$. From the point of view of friction reduction, it is also not wise to dispense very large lubricant volumes upon a pattern decorated with narrow and deep pits.
Therefore, the choice of the lubricant fraction ($a$) is the most significant parameter after the viscosity of the lubricant, which can influence the slip length. 
Ideally, the choice of $100\%$ lubricant fraction provides the maximum possible slip. However, in that case, there will be no rib left that traps the lubricant. This leads to the complete removal of lubricant and the purpose of making LIS vanishes.

\subsection{Anisotropic Effects}   \label{sec:RecRelativeSlip}
As evident from the previous section (Sec. \ref{sec:RecSlipLength}), the slip length is larger in longitudinal flow ($z-$ direction) than in transverse ($x-$ direction). The ratio of these two slip length values $b_{\mid \mid}/b_\perp$ is a concise measure to characterize the anisotropic response to a far-field strain rate applied obliquely to the surface patterning direction, as discussed in the literature \cite{feuillebois2010transverse}.  The anisotropic response corresponds to the velocity at $y=0$ deflect toward the 'fast direction' $z$. This anisotropic surface property holds utility in diverse applications, including particle separation \cite{zhang2012separation, pimponi2014mobility, asmolov2015principles} and mixing \cite{stroock_chaotic_2002, stroock2002patterning, jang2016tensorial}.  In the subsequent discussion, $b_{\mid \mid}/b_\perp$  will be termed ``relative slip length''.

\begin{figure}[htbp]
    \centering
    \begin{subfigure}[t]{0.48\textwidth}
        \centering         
        \includegraphics[width=\textwidth]{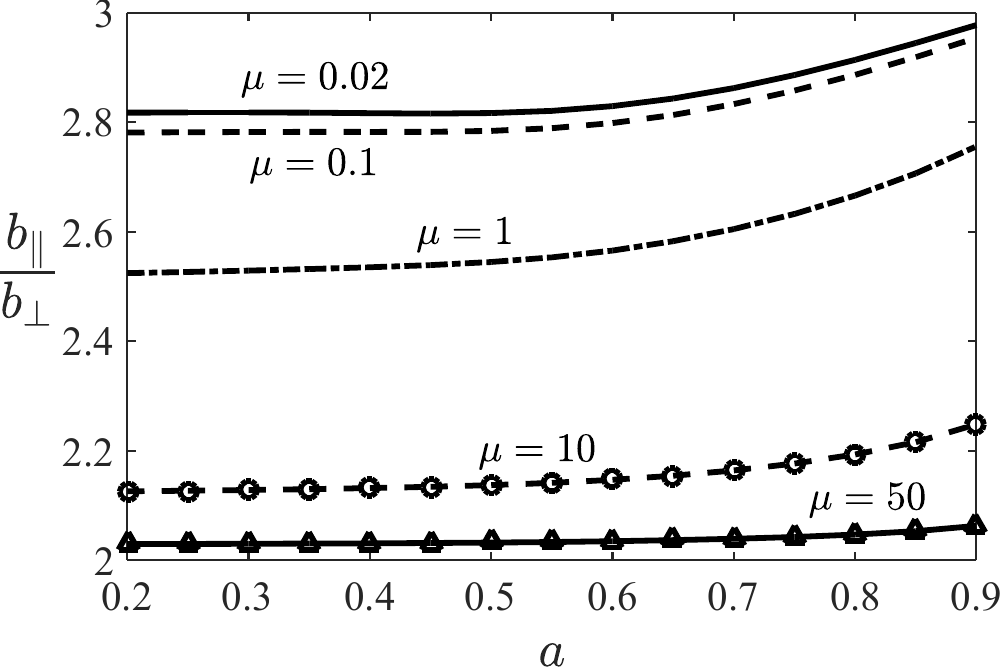}
        \caption{ } 
    \end{subfigure}  
  \caption{Relative slip length variation with lubricant fraction for constant corrugation depth and various main-fluid to lubricant viscosity-ratios.} \label{fig:RecAnisotropicEffect}
\end{figure}

The relative slip length is depicted as a function of lubricant fraction ($a$) across various viscosity ratios ($\mu$) in Fig. \ref{fig:RecAnisotropicEffect}.
At higher viscosity ratios (or lower viscosity of the lubricant), the relative slip length closely approximates the ideal fluid scenario ($b_\parallel/b_\perp = 2$) \cite{asmolov2012effective}, and an increase in the lubricant viscosity results in an increase in the relative slip length. Consequently, increasing the lubricant’s viscosity accentuates the disparity in the slip length magnitude between the direction along the ribs and across it. It is important to highlight that the viscosity ratio dependence of the relative slip length differs from the enhancement in effective slip or friction reduction. While increasing lubricant viscosity consistently diminishes slip, the reduction in slip along the $x-$ direction outpaces the decrease in slip along the $z-$ direction, which ultimately results in the enhancement of the anisotropic surface response with high-viscosity lubricants.

Furthermore, the influence of lubricant fraction on the relative slip length is negligible at high viscosity ratios ($\mu$). However, at lower viscosity ratios, the lubricant fraction ($a$) has a more significant effect, enhancing the relative slip length.

Therefore, it can be inferred that the selection of corrugation shape and lubricant is contingent upon the specific application requirements. For applications like drag reduction, necessitating the maximization of slip, opting for the least viscous lubricant is advisable. Conversely, for applications such as particle separation, where anisotropy in surface properties is beneficial, a higher viscosity lubricant should be preferred to achieve the desired effect.

\section{Conclusion}
In the current study, semi-analytical solutions are obtained for the flow of fluid over lubricant-filled micro-ribbed surfaces. The analytical solution is obtained with no assumption on the fluid's property and rib dimensions. The fluid's interface being flat is the only assumption taken in this study. 
The semi-analytical solutions are obtained using the eigenfunction expansion method for both cases of flow along and across the ribs.

The analytical results are compared with the numerical simulation and literature results for longitudinal and transverse flow, respectively, by comparing the fluid's velocity at the reference plane. 
The current analytical results are found to be accurate for any combination of main and lubricating fluids over any rib dimension.
The present results were also found to be superior to the previously available most accurate theory of the constant stress at the fluid-fluid interface, presented by Sch\"{o}necker et al. \cite{schonecker2014influence}.


The surface-averaged quantity, slip length, is chosen as the metric for analyzing the effect of LIS on the main fluid. 
Using the current analytical results, the effect of fluid property and rib dimensions on the surface's slip length is studied.
The slip length for flow directed along the ribs is always found to be larger than the slip length across the ribs. Moreover, increasing the lubricant's viscosity is also observed to have a detrimental effect on the slip length.

Additionally, regarding the effect of rib dimensions on slip length, it is observed that, between rib height ($d$) and lubricant fraction ($a$), the latter is found to have a larger impact compared to the former. The ratio of slip lengths along the principal directions, termed the relative slip length, is a measure of the anisotropic response of a patterned surface, which can potentially be leveraged for applications such as separations and mixing. It is observed that the anisotropy increases with the decrease in viscosity ratio ($\mu$), underscoring that the choice of the lubricant is highly specific to the desired application.  For example, the same lubricant that works well in reducing the frictional resistance offered by a LIS to a flowing working fluid may be ineffective when used in a particle sorting application based on surface anisotropy  \cite{dubov2017continuous}.

\bibliographystyle{ieeetr}
\bibliography{sample}

\end{document}